\documentstyle[preprint,aps]{revtex}
\title{QCD: from four to two dimensions}
\author{\large A. Bassetto \\[3mm]
\em Dipartimento di Fisica ``G. Galilei'', INFN Sezione di Padova  \\
\em Via Marzolo 8, I-35131 PADOVA (Italy)}
\date{31 August 1998}
\begin{document}
\maketitle
\begin{abstract}
I review some work done in the past four years concerning the 
transition of Yang-Mills theories from 1+3 to 1+1 dimensions.
The problem is considered both in a perturbative context and
in exact solutions when available. Several interesting features
are discussed, mainly in relation to the phenomenon of confinement,
and some controversial issues are clarified. 
\end{abstract}

\vskip 2.0truecm

I would like to report on some work done in the past four years
with the aim of clarifying
properties and peculiarities of a Yang-Mills theory when the
number of space dimensions $d=D-1$ is lowered to $d=1$.

The interest in studying two-dimensional theories is mainly due to 
the possibility of obtaining sometimes exact solutions, which are
believed to share important features with the more realistic 
situation in four dimensions.

Schwinger's model (massless electrodynamics in two dimensions ($QED_2$)) 
is the key example,
which can be exactly solved, exhibiting very
interesting and peculiar properties, like fermion confinement,
theta-vacua and the presence of a non-vanishing chiral condensate.

$QCD_2$ is its non-Abelian generalization and has recently received
most attention in many investigations. It is widely believed that
several phenomena that can be fairly easily understood in two
dimensions, can persist when dimensions are increased.

To be definite, in the following I will limit myself to the ``pure''
Yang-Mills theory ($YM$) with gauge group $SU(N)$ (or sometimes $U(N)$),
in spite of the fact that interesting features emerge when 
dynamical fermions, either in the fundamental or in the
adjoint representation, are present.

I will focus my interest on two topics
\begin{itemize}
\item relations between $YM_4$ and $YM_2$ with a particular care when
considering the limit $D \to 2$;
\item relations between perturbative and non-perturbative solutions.
\end{itemize}

As a technical tool, I will use the Wilson loop, owing to its gauge 
invariance and to its
reasonable infrared (IR) properties. It is indeed well known that,
when approaching $D=2$, ultraviolet (UV) singularities are no longer
a concern, but wild IR behaviours usually show up.

\vskip .5truecm

The investigation started from a perturbative test of gauge invariance
in $YM_4$ at ${\cal O}(g^4)$.

We consider the following Wilson loop
\begin{equation}
{\cal W}_\gamma = {1\over N} \langle 0| {\rm Tr}\left[ {\cal T}{\cal P} 
{\rm exp} \left( ig \oint_\gamma dx^\mu \ A^a_\mu (x) T^a \right)\right]
|0\rangle \ \ , \label{wilson}
\end{equation}
where $\gamma$ is a rectangle with light-like sides parametrized
according to the equations
\begin{eqnarray}
\label{uno}
C_1 &:& x^\mu (t) = n^{* \mu} t,\nonumber\\
C_2 &:& x^\mu (t) = n^{* \mu} + n^\mu t,\nonumber\\
C_3 &:& x^\mu (t) = n^\mu + n^{* \mu}( 1-t), \nonumber\\
C_4 &:& x^\mu (t) = n^\mu (1 - t), \qquad 0 \leq t \leq 1, 
\end{eqnarray}
the vectors $n^\mu=\frac{1}{\sqrt 2}(L,0,0,-L)$ and $n^{* \mu}=
\frac{1}{\sqrt 2}(T,0,0,T)$ 
being indeed light-like and
normalized in such a way that $n\cdot n^*=LT$.

This loop exhibits in four dimensions UV as well as IR singularities;
they were both regularized dimensionally. A calculation in Feynman
gauge was performed in \cite{kk}, whereas in \cite{bkkn} the
same loop was computed in the light-cone gauge $n\cdot A \equiv A_-=0$. 
The two results coincide, as required by gauge invariance, provided
the propagator in light-cone gauge is endowed with a causal
prescription for the ``spurious'' singularity \cite{ml}
$$\frac{1}{n\cdot k}\equiv \frac {1}{n\cdot k+i\varepsilon sign(n^*
\cdot k)}= \frac {n^*\cdot k}{(n\cdot k)(n^*\cdot k)+i\varepsilon},$$
which
naturally follows from canonical equal-time quantization \cite{bdls}.

If instead the Cauchy principal value prescription(CPV) is adopted
$$\frac{1}{n\cdot k}\equiv P(\frac {1}{n\cdot k}),$$ as suggested
by light-front quantization \cite{ks}, causality and thereby 
analyticity properties are 
jeopardized preventing a consistent renormalization of the theory
(see {\it e.g.} \cite{bns}). The result one gets exhibits divercencies
that are not controlled by power counting and does not even resemble
to the one of Feynman gauge.

It is interesting to investigate what happens to the same Wilson loop
calculation when the number $D$ of space-time dimensions approaches 2.
This has been done in ref.\cite{bdg}, both in Feynman and in light-cone
gauge. In Feynman gauge the propagator in strictly 2 dimensions is not
a tempered distribution, owing to its singular IR behaviour. Individual
diagrams, when dimensionally regularized, exhibit poles at $D=2$;
nevertheless these singularities cancel in the sum, leaving a finite
result in the limit $D\to 2$. The same result is recovered in the
light-cone gauge, provided the propagator has the ``spurious'' 
singularity causally prescribed; at $D=2$ it is a tempered
distribution and, indeed, in the coordinate representation
it has the expression
\begin{equation}
D_{++}^{ab}(x) =
{i \delta^{ab}\over \pi^2}\int d^2k
\, e^{ikx} {k_+^2\over (k^2 + i \epsilon)^2}= {\delta^{ab}\over \pi}
{(x^-)^2\over (-x^2 + i\epsilon)}\  \label{propml}
\end{equation}
to be compared with the expression it gets following the CPV prescription
\begin{equation}
D_{++}^{(P)ab}(x)=
- {i\delta^{ab}\over (2\pi)^2} \int d^2k\,
e^{ikx} {\partial\over \partial k_-} P\left({1\over k_-}\right)=-
{i\delta^{ab}\over 2} |x^-|\delta(x^+)\ . \label{propcpv}
\end{equation}
In the light-cone gauge individual diagrams are finite in the limit
$D=2$. Actually the diagram involving the triple vector vertex
vanishes as expected, since in light-cone gauge there is no triple 
vector vertex in two dimensions. Surprisingly, the diagram with
a self-energy loop correction to the vector propagator, does not
vanish; what happens is that the vanishing of triple vertices
is exactly compensated by the loop integration singularity
at $D=2$ leading, eventually, to a finite result. We would like to
stress that it is not a pathology of light-cone gauge; precisely this
term is needed, together with the contribution coming from graphs
with two propagator lines, to get agreement with the Feynman 
gauge result.

According to a general theorem \cite{ft}, the maximally non-Abelian
terms are the relevant ones in perturbative calculations (as a matter
of fact Abelian terms trivially exponentiate); at ${\cal O}(g^4)$
we find
\begin{equation}
{\cal W}^{na}=g^4C_FC_A\frac {{\cal A}^2}{16\pi^2}\bigl(1+\frac {\pi^2}{3}
\bigr),
\label{area}
\end{equation}
the first term coming from the graph containing the self-energy and the
second one from the graph with crossed vector propagators. The quantities
$C_F$ and $C_A$ are the usual quadratic Casimir operators of the fundamental
and of the adjoint representation and ${\cal A}$ is the area of the loop.
 
If the same calculation is performed at exactly $D=2$, the first term is 
obviously missing. Therefore the theory is {\it discontinuous} at $D=2$,
at least in its perturbative formulation.

The occurrence of a term proportional to $C_A$ is troublesome; if indeed 
the same result is obtained for a rectangular loop with sides of length
$2L$ and $2T$,
parallel respectively to a space and to the time directions, this dependence 
would
survive in the large-$T$ limit and would be at odd with the expected
Abelian-like exponentiation in this limit \cite{bns}.

This is the motivation for studying the loop $\gamma$
\begin{eqnarray}
\gamma_1 &:& \gamma_1^\mu (s) = (sT, L)\ ,\nonumber\\
\gamma_2 &:& \gamma_2^\mu (s) = (T,-sL)\ ,\nonumber\\
\gamma_3 &:& \gamma_3^\mu (s) = (-sT, -L)\ , \nonumber\\
\gamma_4 &:& \gamma_4^\mu (s) = (-T, sL)\ , \ \ \qquad -1 \leq s \leq 1. 
\label{path}
\end{eqnarray}
describing a  (counterclockwise-oriented) rectangle
centered at the origin of the plane ($x^1,x^0$),
with sides of length $(2L,2T)$, respectively.

This has been done in two papers \cite{bcn} and \cite{bbn}. In the first one, 
this loop was computed at ${\cal O}(g^4)$ in light-cone gauge in 
exactly 2 dimensions; in the second paper the loop was computed in Feynman
gauge at $D=2+\epsilon$. 

Let us start by discussing the results of the second paper. As long as
$\epsilon>0,$  the loop depends also on the dimensionless ratio
$\beta=\frac{L}{T}$, besides the area. All terms proportional to 
$C_FC_A$ are subleading in the limit $T\to \infty$ with respect to
the ``planar'' terms which are proportional to $C_F^2$. Indeed they
typically behave like $$T^{4-2D}{\cal A}^2.$$

To be more precise \cite{bbn}, the contribution due to diagrams with crossed
propagators, in the large-$T$ limit and for $\omega\equiv D/2$
near 1,  exhibits a double and a single
pole, whose Laurent expansion gives
\begin{equation}
\label{crocilaurent}
{{\cal W}_{na}\pi^{2\omega} e^{2i\pi  \omega}\over g^4 C_FC_A (2T)^{4-4\omega}
 (LT)^2}={1\over 2(\omega-1)^2}+{1-\gamma\over(\omega -1)} -1-2\gamma+\gamma^2
+{\pi^2\over 12} +{\cal O}(\omega -1)\ ,
\end{equation}
$\gamma$ being the Euler-Mascheroni constant.

Similarly, the contribution from diagrams involving the self-energy
corrected vector propagator, is
\begin{equation}
{{\cal W}^{(2)}\pi^{2\omega} e^{2i\pi  \omega}\over g^4 C_FC_A (2T)^{4-4\omega}
 (LT)^2}={1\over (\omega-1)^2}+{9-4\gamma\over2(\omega -1)}+{39\over 2}
-9\gamma +2\gamma^2 +{\pi^2\over 6}
  +{\cal O}(\omega -1)\ ,
\label{bollelaurent}
\end{equation}
and again exhibits a double and a single pole at $\omega=1$.

Finally diagrams involving a triple vector vertex lead to
\begin{eqnarray}
&&\lim_{\beta \to 0}\, {{\cal W}^{(3)}\pi^{2\omega} e^{2i\pi  \omega}
\over g^4 C_FC_A (2T)^{4-4\omega}
 (LT)^2}=-{3\over 2(\omega-1)^2}+{3\gamma -11/2\over(\omega -1)}
-{35\over 2} +11\gamma\nonumber\\
&-&3\gamma^2 +{\pi^2\over 12}
 +{\cal O}(\omega -1)\ .
\label{ragnilaurent}
\end{eqnarray}

Therefore agreement with Abelian-like exponentiation holds and the validity
of previous perturbative tests of gauge invariance in higher
dimensions (see ref.\cite{bns}) 
is fully confirmed. This rather simple and ``universal'' way of
realizing the exponentiation at $D>2$ might have a deeper justification
as well as far-reaching consequences.

However it is clear from the expression above that if we take 
first the limit $\epsilon \to 0$, no damping occurs when $T\to \infty$.
{\it The two limits do not commute}. 

Summing the three contributions, double and single poles at $\omega=1$ 
cancel; when $\epsilon \to 0$
the dependence on $\beta$ disappears and the result of eq.(\ref{area})
is exactly recovered, in spite of the fact that the two loops are
different.

A pure area dependence would be hardly surprising in
view of the invariance of the loop in two dimensions under
area-preserving diffeomorphisms. Still we remind the reader that
the first contribution in eq.(\ref{area}) was obtained after a limiting
procedure from higher dimensions: in exactly two dimensions it does not
occur. Then we find amazing that it respects such a symmetry on its
own.

Finally, in ref. \cite{bcn} the same space-time loop is computed in light-cone
gauge at exactly $D=2$. One obtains only the second term of eq.(\ref{area}),
as expected. 

Let us summarize what we have achieved so far.
\begin{itemize}
\item For $D>2$ the ${\cal O}(g^4)$ result we get
is in agreement with the expected Abelian-like area exponentiation  
in the large-$T$ limit. $C_FC_A$ terms are subleading.
\item In the limit $D\to 2$ the ${\cal O}(g^4)$ result is finite,
it depends only on the area of the loop and consists of two addenda
(see eq.(\ref{area})). No simple area exponentiation occurs in the
large-$T$ limit, owing to the presence of a leading $C_FC_A$ 
contribution. Therefore the limits $T \to \infty$ and $D \to 2$
{\it do not commute}.
\item At exactly $D=2$ only the second term of eq.(\ref{area})
survives. The theory exhibits a discontinity in the limit $D\to 2$
(for any value of $T$!), at least in its perturbative formulation.
\end{itemize}
All the above conclusions are shared by Feynman and light-cone gauges.

Moreover, working at exactly $D=2$, Staudacher and Krauth \cite{sk} were
able to resum in light-cone gauge with causal prescription
the perturbative series at all orders in the coupling constant $g$, 
thereby generalizing our ${\cal O}(g^4)$ result. They get for $U(N)$
in the Euclidean formulation
\begin{equation}
\label{krauth}
{\cal W}=\frac{1}{N}exp\Big[- \frac { g^2\cal A}{4}\Big]L^1_{N-1}
\Bigl(\frac{g^2 {\cal A}}{2}\Bigr),
\end{equation}
the function $L^1_{N-1}$ being a generalized Laguerre polynomial.

This result is definitely {\it different} from the exact expression
which is known in two dimensions
\begin{equation}
\label{exact}
{\cal W}= exp\Big[-g^2 \frac {N \cal A}{4}\Big],
\end{equation}
and has been obtained by different authors using different procedures.

Not only a Laguerre polynomial appears as a factor in eq.(\ref{krauth}), 
but also the
string tension, namely the constant in the exponential, turns out to
be different from the expected one.

More dramatically, eq.(\ref{krauth}) in the limit $N\to \infty$ with
$\hat g^2=\frac{g^2 N}{2}$ fixed, becomes
\begin{equation}
\label{bessel}
{\cal W}\to \frac{1}{\sqrt{\hat g^2 {\cal A}}}J_1(2\sqrt{\hat g^2 {\cal A}}),
\end{equation}
and confinement is lost.

So what is wrong (if anything) with eq.(\ref{krauth})?

In order to understand this point, it is worthwhile to study the problem
on a compact two-dimensional manifold, that, for simplicity, we choose
the sphere $S^2$ \cite{bg}. We shall also consider the slightly simpler
case of the group $U(N)$ (the generalization to $SU(N)$ is straightforward).  
On $S^2$ we consider a smooth non self-intersecting closed contour 
$\Gamma$.
We call $A$ the total area of the sphere,
which eventually will be sent to $\infty$, whereas ${\cal A}$ will be
the area ``inside'' the loop we keep finite in this limit.
It is well known that, on $S^2$ at large N, a phase transition occurs
between two regimes, a weak coupling regime which correspond to small
values of $g^2 A$ and a strong coupling regime for large $g^2 A$
\cite{dg}. This phase transition is driven by instantons.
We follow closely the treatment of this problem given in refs.\cite{bg},
\cite{gm}.

Our starting point are the well-known expressions \cite{Blau} of the
exact partition function and of a non self-intersecting Wilson loop
for a pure $U(N)$ Yang-Mills theory on a sphere $S^2$ with area $A$
\begin{equation}
\label{partition}
{\cal Z}(A)=\sum_{R} (d_{R})^2 \exp\left[-{{g^2 A}\over 4}C_2(R)\right],
\end{equation}
\begin{eqnarray}
\label{wilsonz}
{\cal W}(A,{\cal A})&=&{1\over {\cal Z}N}\sum_{R,S} d_{R}d_{S}
\exp\left[-{{g^2 {\cal A}}\over 4}C_2(R)-{{g^2 (A-{\cal A})}\over 4}
C_2(S)\right]\nonumber \\
&&\times \int dU {\rm Tr}[U]\chi_{R}(U) \chi_{S}^{\dagger}(U),
\end{eqnarray}
$d_{R\,(S)}$ being the dimension of the irreducible representation $R(S)$ of
$U(N)$; $C_2(R)$ ($C_2(S)$) is the quadratic Casimir, 
the integral in (\ref{wilsonz}) is over the
$U(N)$ group manifold while $\chi_{R(S)}$ is the character of the group
element $U$ in the $R\,(S)$ representation. 
From eq.(\ref{wilsonz}) it is possible to derive, in the large-$A$ 
decompactification limit,
the behaviour (\ref{exact})\cite{boul}.

We write these equations explicitly for $N>1$ in the form
\begin{eqnarray}
\label{partip}
{\cal Z}(A)&=&\frac{1}{N!}\exp \left [ -\frac{g^2 A}{48}N(N^2-1)\right]
\nonumber \\
&&\times \sum_{m_i = -\infty}^{+\infty}\Delta^2(m_1,...,m_N)\exp\left [ 
-\frac{g^2A}{4}\sum_{i=1}^N (m_i-\frac{N-1}{2})^2\right],
\end{eqnarray}
\begin{eqnarray}
\label{wilsonp}
&&{\cal W}(A,{\cal A})=\frac{1}{{\cal Z}N}\exp \left[-\frac{g^2A}{48}N(N^2-1)
\right ]\frac{1}{N!}\nonumber \\
&& \times \sum_{k= 1}^{N}\sum_{m_i=-\infty}^{+\infty}
\Delta(m_1,...,m_N)\Delta(m_1+\delta_{1,k},...,m_N+\delta_{N,k})\nonumber\\ 
&&\times \exp\left [-\frac{g^2 {\cal A}}{4}\sum_{i=1}^N (m_i-\frac{N-1}{2})^2
 -\frac{g^2 (A-{\cal A})}{4}\sum_{i=1}^N (m_i-\frac{N-1}{2}+\delta_{i,k})^2\right].
\end{eqnarray}
We have described the generic irreducible representation by means
of the set of integers $m_{i}=(m_1,...,m_{N})$, related to the
Young tableaux, in terms of which
we get
\begin{eqnarray}
\label{casimiri}
C_2(R)&=&\frac{N}{12}(N^2-1)+\sum_{i=1}^{N}(m_{i}-\frac{N-1}{2})^2,\nonumber
\\
d_{R}&=&\Delta(m_1,...,m_{N}).
\end{eqnarray}
$\Delta$ is the Van der Monde determinant and
 the integration in eq.(\ref{wilsonz})
has been performed explicitly, using the well-known formula for the 
characters in terms of the set $m_{i}$.

Now, as first noted by Witten \cite{Witte}, it is possible to
represent ${\cal Z}(A)$ (and consequently ${\cal W}(A,{\cal A})$) as a sum over
instable instantons, where each instanton contribution is 
associated to a finite,
but not trivial, perturbative expansion. The easiest way to see it, is 
to perform a Poisson resummation in eqs. (\ref{partip}),(\ref{wilsonp})
\cite{Poli}

\begin{eqnarray}
\label{poisson}
&&\sum_{m_{i}=-\infty}^{+\infty}F(m_1,...,m_{N})=
\sum_{n_{i}=-\infty}^{+\infty}\tilde{F}(n_1,...,n_{N}),\nonumber\\
&&\tilde{F}(n_1,...,n_{N})=\int_{-\infty}^{+\infty}dz_1...dz_{N}
\exp \left[2\pi i(z_1 n_1+...+z_{N}n_{N})\right]F(z_1,...,z_{N}).
\end{eqnarray}

We have carefully repeated the original computations of ref.\cite{gm},
paying particular attention to the numerical factors and to the area
dependences; as a matter of fact, at variance with \cite{gm}, 
where interest
was focussed on the large-$N$ limit, we are mainly concerned with 
decompactification (large $A$) and with a comparison with the results
of ref.\cite{sk} for any value of $N$. We have obtained
\begin{eqnarray}
\label{istanti}
{\cal Z}(A)&=&C(g^2 A,N)\sum_{n_{i}=-\infty}^{+\infty}
\exp\left[-S_{inst}(n_{i})\right]Z(n_1,...,n_{N}),\nonumber\\
{\cal W}(A,{\cal A})&=&\frac{1}{{\cal Z}N}C(g^2 A,N)\exp \left[
-g^2\frac{{\cal A}(A-{\cal A})}{4A}\right]\sum_{n_{i}=-\infty}^{+\infty}
\exp\left[-S_{inst}(n_{i})\right]\nonumber\\
&\times&\sum_{k=1}^{N}\exp\left[-2 \pi i n_{k}\frac{A-{\cal A}}{A}\right]
W_{k}(n_1,...,n_{N}),
\end{eqnarray}
where
\begin{eqnarray}
\label{quantita`}
C(g^2 A,N)&=& (i)^{N(N-1)}{{{\frac{g^2 A}{2}}^{-N^2/2}}\over {N!}}\exp
\left[-\frac{g^2 A}{48}N(N^2-1)\right]\nonumber\\
S_{inst}(n_{i})&=&\frac{4\pi^2}{g^2 A}\sum_{i=1}^{N}n_{i}^2,
\end{eqnarray}
and
\begin{eqnarray}
\label{zetawu}
Z(n_1,...,n_{N})&=&\exp(i\pi (N-1)\sum_{i=1}^{N}n_{i})\int
_{-\infty}^{+\infty}dz_1...dz_{N}\exp\left[-\frac{1}{2}
\sum_{i=1}^{N}z_{i}^2\right]\nonumber\\
&\times&\prod_{i<j}^{N}\Big(\frac
{8 \pi^2}{g^2 A}(n_{i}-n_{j})^2-(z_{i}-z_{j})^2\Big),\nonumber\\
W_{k}(n_1,...,n_{N})&=&\exp(i\pi (N-1)\sum_{i=1}^{N}n_{i})\int
_{-\infty}^{+\infty}dz_1...dz_{N}\exp\left[-\frac{1}{2}
\sum_{i=1}^{N}z_{i}^2\right]\times\\
\prod_{i<j}^{N}\Bigl[\Big(\frac
{2\sqrt{2} \pi}{\sqrt{g^2 A}}(n_{i}-n_{j})&+&ig^2\frac{A-2{\cal A}}
{\sqrt{2g^2 A}}
(\delta_{i,k}-\delta_{j,k})\Big)^2
-\Big((z_{i}-z_{j})+i\frac{\sqrt{g^2 A}}{2\sqrt 2}
(\delta_{i,k}-\delta_{j,k})\Big)
^2\Bigr].\nonumber
\end{eqnarray}

These formulae have a nice interpretation in terms of instantons.
Indeed, on $S^2$, there are non trivial solutions of the Yang-Mills equation,
labelled by the set of integers $n_{i}=(n_1,...,n_{N})$ 
\begin{equation}
\label{monopolo}
{\cal A}_{\mu}(x)=\left(\matrix{n_1 {\cal A}_{\mu}^{0}(x) & 0 & \ldots & 0 \cr
                         0      & n_2 {\cal A}_{\mu}^{0}(x) & \ldots &0\cr
                         0 &\ldots&\ldots&0\cr
                         \vdots&\vdots&\ddots&\vdots\cr
                         0& 0 &\ldots &n_N{\cal A}_{\mu}^{0}(x)\cr
}\right)
\end{equation}
where ${\cal A}_{\mu}^{0}(x)={\cal A}_{\mu}^{0}(\theta, \phi)$ is the Dirac
monopole potential,
$${\cal A}_{\theta}^{0}(\theta, \phi)=0 , \quad\   {\cal A}_{\phi}^{0}
(\theta, \phi)={1-\cos \theta\over 2}, $$
$\theta$ and $\phi $ being the polar (spherical) coordinates on $S^{2}$.

From the above representations it is rather clear why the decompactification
limit $A\to \infty$ should not be performed too early. Indeed on the plane
it is not easy to distinguish fluctuations around the instanton solutions
from Gaussian fluctuations around the trivial field configuration, since
$S_{inst}(n_{i})$ goes to zero for any finite set $n_{i}$ when $A\to
\infty$. For finite $A$ and finite $n_{i}$ instead, 
in the limit $g\to 0$, only the zero
instanton sector can survive in the Wilson loop expression (notice that
the power-like singularity $(g^2)^{-N^2/2}$ in the coefficient $C(g^2 A,N)$
exactly cancels in the normalization). In this limit each instanton 
contribution is ${\cal O}(\exp(-\frac{1}{g^2}))$; therefore instantons 
become crucial only when they are completely resummed.

On the other hand the zero instanton contribution should be obtainable 
in principle 
by means of perturbative calculations.  

In the following we compute from eqs.(\ref{istanti}) the exact
expression on the sphere $S^2$  of the zero instanton contribution
to the Wilson loop, obviously normalized to zero instanton 
partition function.

We write eq.(\ref{istanti}) for the zero instanton sector $n_{i}=0$.
Thanks to its symmetry, we can always choose $k=1$ and the equation becomes
\begin{eqnarray}
\label{zeroinst}
{\cal W}_0 &=&(2\pi)^{-\frac{N}{2}}\prod_{n=0}^{N}\frac{1}{n!}
\exp\left[-g^2\frac{{\cal A}(A-{\cal A})}{4A}\right]
\int_{-\infty}^{+\infty}dz_1...dz_{N}\exp\left[-\frac{1}{2}\sum_{i=1}^{N}
z_{i}^2\right]\nonumber\\
&\times&\prod_{j=2}^{N}\Big[(z_1-z_{j})^2+i\sqrt{{g^2A}\over 2}(z_1-z_{j})-
g^2\frac{{\cal A}(A-{\cal A})}{2A}\Big]\Delta^2(z_2,...,z_{N}).
\end{eqnarray}
We introduce the two roots of the quadratic expression in the integrand
$z_{\pm}=z_1+i\alpha\pm i\beta$ with $\alpha=\frac{\sqrt{g^2 A}}{2\sqrt 2}$
and $\beta=\frac{\sqrt{g^2}(2{\cal A}-A)}{2\sqrt{2A}}$.
The previous equation then becomes
\begin{eqnarray}
\label{zetapiu}
{\cal W}_0 &=&(2\pi)^{-\frac{N}{2}}\prod_{n=0}^{N}\frac{1}{n!}
\exp\left[-g^2\frac{{\cal A}(A-{\cal A})}{4A}\right]
\int_{-\infty}^{+\infty}dz_1...dz_{N}\exp\left[-\frac{1}{2}\sum_{i=1}^{N}
z_{i}^2\right]\nonumber\\
&\times&\Delta(z_{+},z_2,...,z_{N})\Delta(z_{-},z_2,...,z_{N}).
\end{eqnarray}
The two Van der Monde determinants can be expressed in terms of
Hermite polynomials \cite{gm} and then expanded in the usual way.
The integrations over $z_2,...,z_{N}$ can be performed, taking
the orthogonality condition into account; we get 
\begin{eqnarray}
\label{integrata}
{\cal W}_0 &=&(2\pi)^{-\frac{1}{2}}\prod_{n=0}^{N}\frac{1}{n!}
\exp\left[-g^2\frac{{\cal A}(A-{\cal A})}{4A}\right]
\prod_{k=2}^{N}(j_{k}-1)! 
\varepsilon^{j_1...j_{N}}\varepsilon_{j_1...j_{N}}\nonumber\\
&\times&\int_{-\infty}^{+\infty}dz_1\exp\Big[-\frac{z_1^2}{2}
\Big]He_{j_1-1}(z_{+})He_{j_1-1}(z_{-}).
\end{eqnarray}
Thanks to the relation
\begin{equation}
\label{laguerre}
\int_{-\infty}^{+\infty}dz_1\exp\Big[-\frac{z_1^2}{2}\Big]
He_{j_1-1}(z_{+})He_{j_1-1}(z_{-})=\sqrt{2\pi}(j_1-1)! L_{j_1-1}
(\alpha^2-\beta^2),
\end{equation}
we finally obtain our main result
\begin{equation}
\label{risultato}
{\cal W}_0=\frac{1}{N}\exp\left[-g^2\frac{{\cal A}(A-{\cal A})}{4A}\right]
\, L_{N-1}^1(g^2\frac{{\cal A}(A-{\cal A})}
{2A}).
\end{equation}

At this point we remark that, in the decompactification limit $A\to
\infty, {\cal A}$ fixed, the quantity in the equation above {\it exactly}
coincides, for any value of $N$, with eq.(\ref{krauth}), which
was derived following completely different considerations.
We recall indeed that that result was obtained by a full resummation
at all orders of the perturbative expansion of the Wilson loop
in terms of Yang-Mills propagators in light-cone gauge,
endowed with the causal prescription.

We conclude that,
for any value of $N$, the pure area law exponentiation (eq.(\ref{exact})) 
follows, after
decompactification, only by
resumming all instanton sectors, a procedure
which changes completely the zero
sector behaviour and, in particular, the value of the string tension.

In the light of the considerations above, there is no contradiction between
the use of the causal prescription in the light-cone propagator and the pure
area law exponentiation of eq.(\ref{exact}); 
this prescription is correct but the ensuing
perturbative calculation can only provide us with the expression 
for ${\cal W}_0$. The paradox of ref.\cite{sk} is solved
by recognizing that they did not take into account the genuine ${\cal
O}(\exp(-\frac{1}{g^2}))$ non
perturbative quantities coming, after decompactification, from the
instantons on the sphere.

We find quite remarkable that {\it both} expressions in eqs.(\ref{krauth})
and (\ref{exact}), respectively, are (different) {\it analytic} 
functions of $g^2$. This is hardly surprising for eq.(\ref{krauth}), but
not for eq.(\ref{exact}), if it is thought as a sum over instanton 
contributions. This analytic behaviour is at the root of the possibility
of obtaining eq.(\ref{exact}) in a quite different way.
As a matter of fact, if the Wilson loops 
we have previously considered, are computed using the 
istantaneous 't Hooft-CPV potential of eq.(\ref{propcpv}), 
which follows from
light-front quantization, and just 
resumming at all orders the
related perturbative series, one exactly recovers the
correct pure area exponentiation (\ref{exact}), {\it i.e.} the same result 
which
requires the essential introduction of non perturbative effects (intantons),
when studied in equal-time quantization \cite{bcn}.

This surprising feature is the origin of almost all recent calculations 
in $QCD_2$,
which make essential use of the 't Hooft's propagator (\ref{propcpv}). 

Confinement in this
picture occurs as a trivial generalization of the $QED_2$ situation,
namely as a consequence of a two
dimensional ``instantaneous'' increasing potential between a $q\bar q$ pair,
giving rise to hadronic strings in a natural way (we recall that, with the
potential (\ref{propcpv}), only planar graphs survive and therefore
only $C_F$ can appear). But we feel unlikely that a similar
mechanism can be at the root of a realistic confinement in higher dimensions.

A deeper conjecture, which deserves further study, may relate it
to some peculiar properties of the light-front vacuum
(we remind the reader that the light-cone CPV prescription follows
from canonical light-front quantization \cite{bns}). In equal-time
quantization ``axial'' ghosts are present \cite{bns},\cite{bdg},
which, although expunged from the ``physical'' Hilbert space,
contribute to Green functions. These degrees of freedom are
canonically suppressed when quantizing on the light-front. In two dimensions
this procedure might perhaps be viable in a ``continuum''
formulation, as renormalization is no longer a concern, 
but, in higher dimensions, it is certainly illegitimate and perturbatively
incorrect. It can neither be extended smoothly beyond the strictly
two-dimensional case nor it can be smoothly
continued to any Euclidean formulation and compared to 
different gauge choices.

Why the instantons we have hitherto considered 
seem to be crucial only in two dimensions in order 
to obtain the correct
area exponentiation, is at present an
interesting open problem.

\end{document}